\documentclass{ws-ijmpcs}

\begin{document}

\markboth{K.~P.~Khemchandani, A.~Mart\'inez~Torres, H.~Nagahiro, A.~Hosaka}
{$N^*$'s and $\Delta$'s generated in vector, pseudoscalar meson-baryon systems}

%
\catchline{}{}{}{}{}
%

\title{$N^*$'s AND $\Delta$'s GENERATED IN VECTOR, PSEUDOSCALAR MESON-BARYON SYSTEMS}

\author{K.~P.~Khemchandani}

\address{Instituto de F\'isica, Universidade de S\~ao Paulo, C.P 66318, 05314-970 S\~ao Paulo, SP, Brazil \\
kanchan@if.usp.br}

\author{A.~Mart\'inez~Torres}

\address{Instituto de F\'isica, Universidade de S\~ao Paulo
C.P 66318, 05314-970 S\~ao Paulo, SP, Brazil\\
amartine@if.usp.br}

\author{H.~Nagahiro}

\address{Department of Physics, Nara Women's University,  Nara 630-8506, Japan\\
and\\
Research Center for Nuclear Physics (RCNP), Mihogaoka 10-1, Ibaraki 567-0047, Japan.\\
nagahiro@rcnp.osaka-u.ac.jp}

\author{A.~Hosaka}

\address{
Research Center for Nuclear Physics (RCNP), Mihogaoka 10-1, Ibaraki 567-0047, Japan.\\
hosaka@rcnp.osaka-u.ac.jp}
\maketitle

\begin{history}
\received{Day Month Year}
\revised{Day Month Year}
\end{history}

\begin{abstract}
We have recently studied coupled channel  interactions of  vector and pseudoscalar mesons with octet baryons having total strangeness null and spin-parity $1/2^-$. We, thus, considered $\pi N$, $\eta N$, $K \Lambda$, $K \Sigma$ , $\rho N$, $\omega N$, $\phi N$, $K^* \Lambda$, and $K^*\Sigma$ with total isospin 1/2 and 3/2. 
The interactions between  pseudoscalar mesons and baryons are obtained by using the Weinberg-Tomozawa theorem. For the case of vector mesons, we calculate $s$-, $t$-, $u$-exchange diagrams and a contact term. The subtraction constants, required to calculate the loop-function in the scattering equations, are constrained by fitting the available experimental data on some of the reactions with pseudoscalar meson-baryon  final states. We end up  finding  resonances which can be related to  $N^*(1535)$, $N^*(1650)$ (with a double pole structure),  $N^*(1895)$ and $\Delta(1620)$.  We conclude that these resonances can be, at least partly, interpreted as dynamically generated resonances and that  the vector mesons play an important role in determining the dynamical origin of the low-lying  $N^*$ and $\Delta$ states.

\keywords{$N^*$-resonances, Bethe-Salpeter equations, Hidden local symmetry.}
\end{abstract}

\ccode{PACS numbers: 13.75.-n, 14.20.Gk}

\section{Introduction}	
In this manuscript we outline some of the important features of our recent work on the nonstrange meson-baryon systems~\cite{us_recent}. The basis of this work lies in the framework built in our previous studies ~\cite{us1,us2} by using effective field theories relying on the chiral and hidden local symmetries (HLS)~\cite{bando} . We first investigated  the reliability of low energy theorems when dealing with vector meson-baryon (VB) systems in Ref.~\refcite{us1} and found that the diagrams other than the $t$-channel  give significant contributions and  it is important to consider $s$-, and $u$-channel diagrams. In addition to these, a contact term  arises from the VB Lagrangian which, apart from its large contribution, is important to be considered from the point of view of the HLS gauge invariance~\cite{us1}.

Having resolved this issue, we included pseudoscalar mesons in the formalism~\cite{us2} and studied the strangeness $-1$ sector  where we found dynamical generation of some hyperon resonances. In continuation with these works, we now deal with strangeness 0 meson baryon systems. Several studies have been dedicated to understand the nature of $N^*$ and $\Delta$ resonances  (see, e.g., Ref~\refcite{juelich,lutz,vlc_group,jujun,3b}). The picture, however,  seems to remain unclear~\cite{us_recent}. We explore these resonances within our formalism, which is briefly discussed in the next section, and find that some $N^*$'s and $\Delta$'s can be partly understood as dynamically generated resonances with the vector meson-baryon contribution being an important one~\cite{us_recent} .

\section{Formalism}
The general VB Lagrangian in our formalism is  written as~\cite{us1} 
\begin{eqnarray} \label{vbb}\nonumber
\mathcal{L}_{\rm VB}&=& -g \Biggl\{ \langle \bar{B} \gamma_\mu \left[ V_8^\mu, B \right] \rangle + \langle \bar{B} \gamma_\mu B \rangle  \langle  V_0^\mu \rangle  +  \frac{1}{4 M} \biggl( F \langle \bar{B} \sigma_{\mu\nu} \left[ V_8^{\mu\nu}, B \right] \rangle\biggr.
\Biggr. \\ \label{eq}
&+&\left. \biggl. D \langle \bar{B} \sigma_{\mu\nu} \left\{ V_8^{\mu\nu}, B \right\} \rangle +  C_0  \langle \bar{B} \sigma_{\mu\nu}  V_0^{\mu\nu} B  \rangle \biggr) \right\},
\end{eqnarray}
where $V$ with subscript $8$ ($0$)  represents the octet (singlet) vector field, the constants $D= 2.4, F= 0.82$  and $C_0 = 3F-D$, such  that the anomalous magnetic couplings: $\kappa_\rho \simeq$ 3.2,  $\kappa_\omega \simeq \kappa_\phi =$ 0. As discussed above, the VB interaction is obtained from the  $s$-, $t$- and $u$-channel diagrams and a contact term.  Using Eq.~(\ref{vbb}) we obtain the Yukawa type vertices necessary to calculate different exchange diagrams, while the contact term arises from the commutator  of the vector meson tensor ($V^{\mu\nu}$).

Next,  the pseudoscalar mesons are included by extending the Kroll-Ruderman theorem for the photoproduction of pions by replacing the photon by a vector meson which is introduced as the gauge boson of the HLS. The  transition amplitudes for VB $\to$ pseudoscalar meson-baryon (PB) are obtained from~\cite{us2}
\begin{eqnarray}
\mathcal{L}_{\rm PBVB} = \frac{-i g_{KR}}{2 f_\pi} \left ( \tilde{F} \langle \bar{B} \gamma_\mu \gamma_5 \left[ \left[ P, V^\mu \right], B \right] \rangle + 
\tilde{D} \langle \bar{B} \gamma_\mu \gamma_5 \left\{ \left[ P, V^\mu \right], B \right\}  \rangle \right), \label{pbvb}
\end{eqnarray}
where $\tilde{F} = 0.46$, $\tilde{D}=0.8$ such that  $\tilde{F} + \tilde{D} \simeq  g_A = 1.26$.

 Finally,  we rely on the Weinberg-Tomozawa theorem for the PB interaction.

\section{Results and Summary}
With the interaction kernels obtained by using the Lagrangians given in the previous section, we solve the Bethe-Salpeter equation in the coupled channel approach.
The loop functions for the same are calculated using the dimensional regularization scheme, the subtraction constants required for  which are fixed by fitting the data on the   isospin 1/2 and 3/2  $\pi N$amplitudes and on the reactions: $\pi^- p \to \eta n$  and $\pi^- p \to K^0 \Lambda$. The best fit is obtained for the subtraction constants given in Table~\ref{afit}. 
\begin{table}[h!]
\tbl{The  subtraction constants required to calculate the loop functions.}
{\begin{tabular}{@{}ccccccccc@{}} \toprule
$\pi N$ &$\eta N$  &$K \Lambda$ &$K \Sigma$& $\rho N$&  $\omega N$& $\phi N$ &$K^* \Lambda$&$K^* \Sigma$ \\\hline
 -1.955& -0.777& -4.476  & -1.945 & -0.45  & -0.955 & -2.972 & -0.184& -1.152    \\ \botrule
\end{tabular} \label{afit}}
\end{table}
These constants are such that the loop functions below the threshold (at least near the threshold region) are real and negative valued, which partly satisfies the condition (of the natural renormalization scheme of Ref.~\refcite{hyodo} ) required to interpret the resonances found in these systems as dynamically generated ones. By saying partly, we mean that the subtraction constants, although being negative, are  different  from those obtained in this natural scheme (see detailed discussions in Ref.~\refcite{us_recent}). This implies that there is something missing in the present formalism (which could be the quark components, other hadron channels, other diagrams, etc.) which is compensated by the difference in the subtraction constants used in our work and those obtained in the natural scheme. 
In summary, we can interpret the resonances found in our work as partly dynamically generated ones.

\begin{figure}[h!]
\centerline{\psfig{file=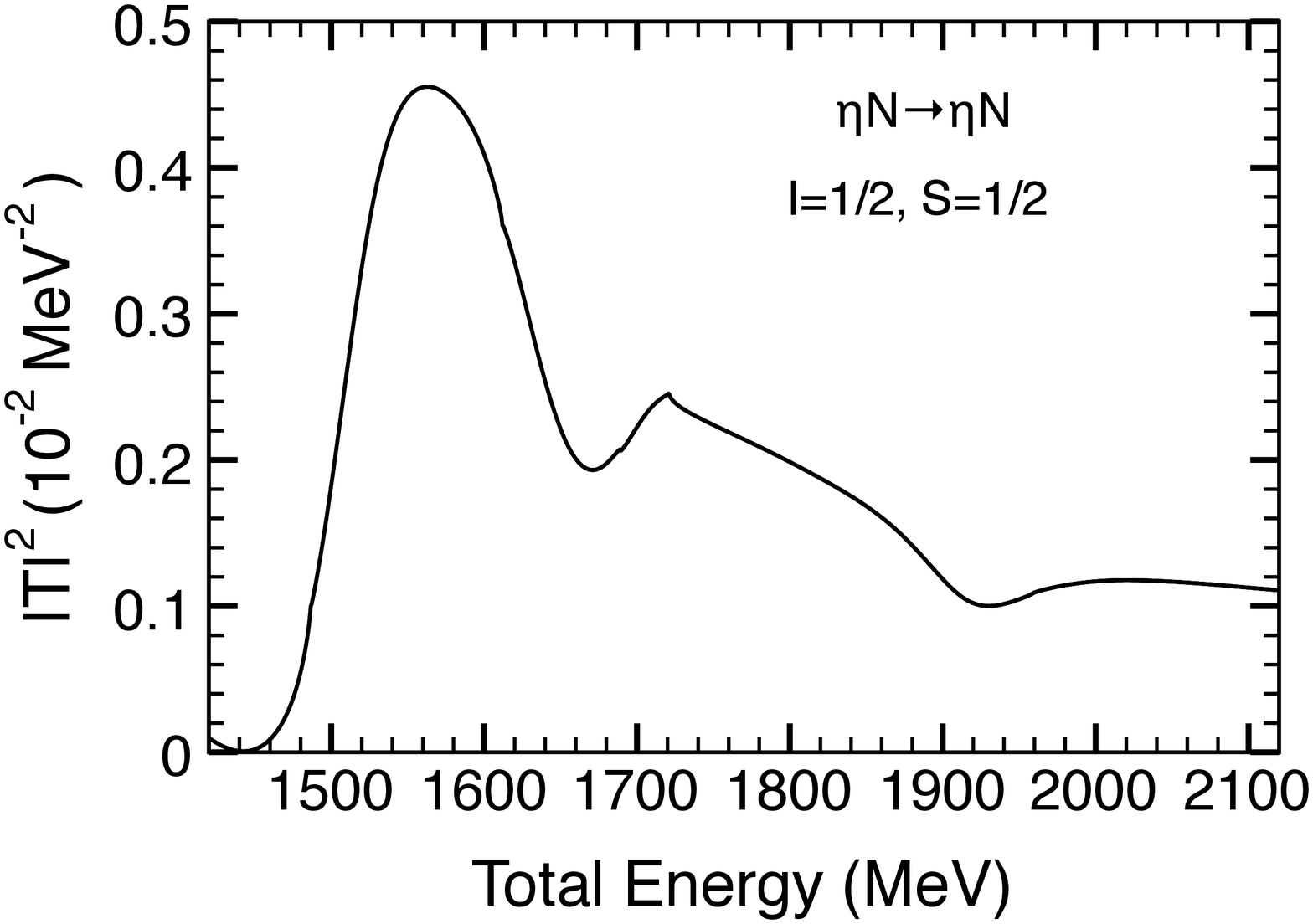,width=5.5cm}
\psfig{file=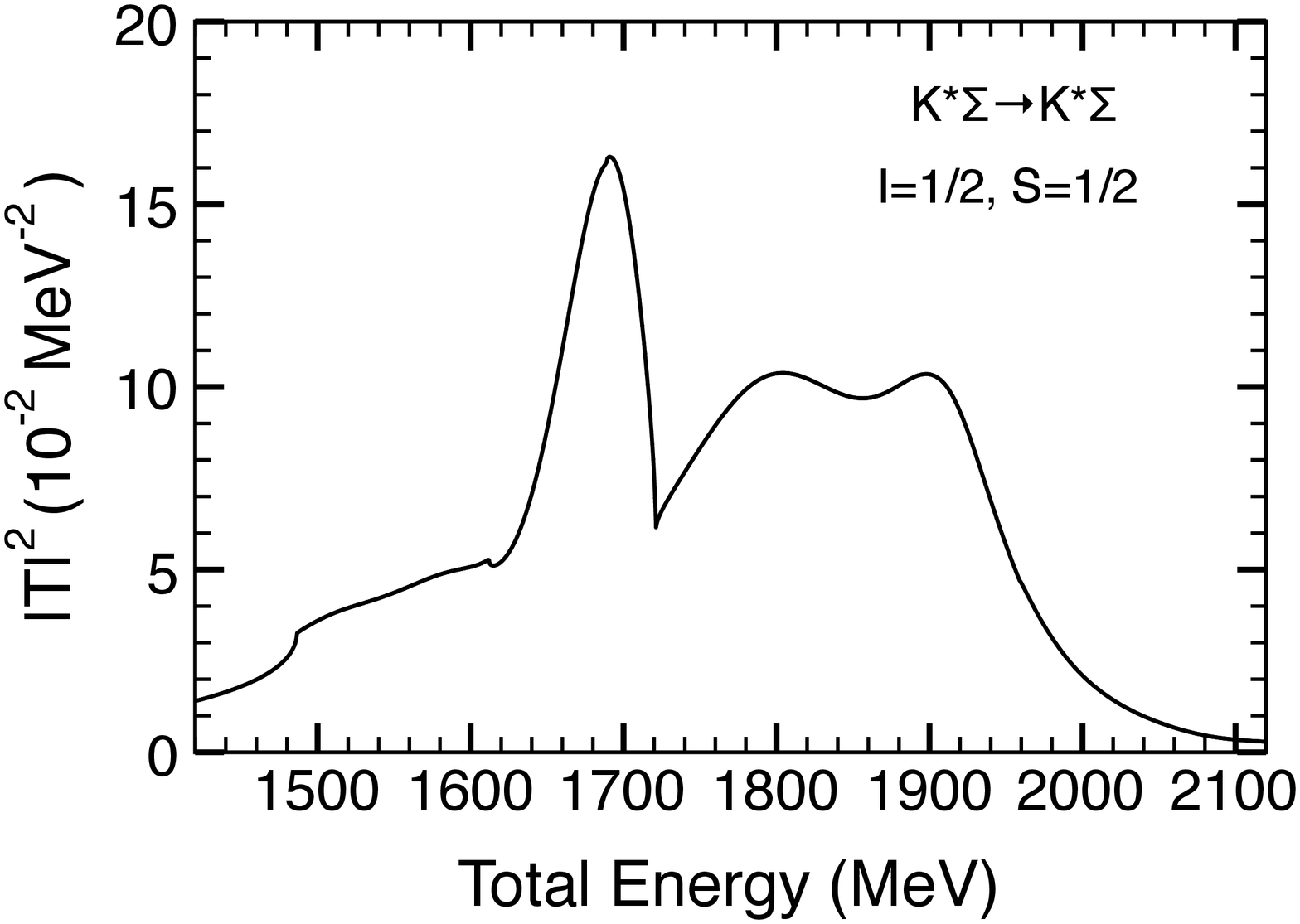,width=5.5cm}}
\caption{Squared amplitudes for the $\eta N$ and $K^*\Sigma$ channels. 
\label{f1}}
\end{figure}

We show the squared amplitudes for the $\eta N$ and the $K^* \Sigma$ channel in the isospin 1/2 configuration in Fig.~\ref{f1}.
 As can be seen from these figures, the $\eta N$ amplitude shows a clear peak structure for $N^*(1535)$ while the $K^*\Sigma$ amplitude shows a peak structure which can be identified with $N^*(1650)$ and a double hump which we relate to $N^*(1895)$. We have studied these amplitudes in the complex plane too and found a pole at $1504 - i55$
MeV  which decays with the branching ratio of 43$\%$  to the $\pi N$  and 55$\%$ to the $\eta N$ channel. All these findings are in good agreement with the properties of the $N^*(1535)$ resonance.  In case of $N^*(1650)$, we find a double pole structure associated to it (with $M_R - i\Gamma/2 =1668 - i28, 1673 - i67$), like in Ref.~\refcite{arndt}. Beyond this energy region
we find two poles at: $1801 - i96$ MeV and  $1912 - i54$ MeV.  All the results for a possible third isospin 1/2 resonance with spin-parity $1/2^-$ are grouped under $N^*(1895)$ in the particle data book~\cite{pdg}. Our results show that there might be two resonances beyond $N^*(1650)$ but there large overlapping widths may not allow to distinguish them in the experimental data.

Finally, we would like to add that we find a pole in the isospin 3/2 configuration  which can be related to $\Delta (1600)$.

\section*{Acknowledgments}
 K.~P.~K  and A.~M.~T thank the Brazilian funding agencies FAPESP and CNPq.
This work is partly  supported  by the Grant-in-Aid for Scientific Research on Priority Areas titled ÒElucidation of New Hadrons with
a Variety of Flavors" (E01: 21105006 for  A.~H) and (24105707 for H.~N) and the authors acknowledge the same.


\end{document}